\input harvmac
\noblackbox
\font\tenbifull=cmmib10 
\font\tenbimed=cmmib10 scaled 800
\font\tenbismall=cmmib10 scaled 666
\textfont9=\tenbifull \scriptfont9=\tenbimed
\scriptscriptfont9=\tenbismall

\def\IZ{\relax\ifmmode\mathchoice
{\hbox{\cmss Z\kern-.4em Z}}{\hbox{\cmss Z\kern-.4em Z}}
{\lower.9pt\hbox{\cmsss Z\kern-.4em Z}}
{\lower1.2pt\hbox{\cmsss Z\kern-.4em Z}}\else{\cmss Z\kern-.4em Z}\fi}
\def\M{{\cal M}}
\def\valpha{{\fam=9{\mathchar"710B } }}

\font\ticp=cmcsc10
\font\cmss=cmss10 \font\cmsss=cmss10 at 7pt

\font\ticp=cmcsc10
\font\ttsmall=cmtt10 at 8pt

%
%

\def\NP{{\it Nucl. Phys.\ }}

\def\PL{{\it Phys. Lett.\ }}
\def\PR{{\it Phys. Rev.\ }}
\def\PRL{{\it Phys. Rev. Lett.\ }}

\def\JTPL{{\it JETP Lett.\ }}

%

%

%


\def\({\left (}
\def\){\right )}
%
\lref\sen{A. Sen, hep-th/9402032,
\PL {\bf B329} (1994) 217.}
\lref\mont{C. Montonen and D. Olive, \PL {\bf 72B} (1977) 117.}
\lref\osborn{H. Osborn, \PL {\bf 83B} (1979) 321.}
\lref\atiyah{M. F. Atiyah and N. J. Hitchin, ``The Geometry and Dynamic of
Magnetic Monopoles,'' Princeton Univ. Press, Princeton, NJ, 1988.}
\lref\hooft{G. 't Hooft, \NP {\bf B79} (1974) 276;
A.M. Polyakov, \JTPL {\bf 20} (1974) 194.}
\lref\zee{B. Julia and A. Zee, \PR {\bf D11} (1975) 2227.}
\lref\joe{J. Polchinski, hep-th/9510017, \PRL {\bf 75} (1995) 4724; S.
Chaudhuri, C. Johnson and J. Polchinski, hep-th/9602052.}
\lref\witten{E. Witten, hep-th/9507121.}
\lref\witdy{E. Witten, hep-th/9503124, \NP {\bf B443} (1995) 85.}
\lref\aspin{P. Aspinwall, hep-th/9507012, \PL {\bf B357} (1995) 329.}
\lref\duff{M.J. Duff and J.X. Lu, \PL {\bf B273} (1991) 409.}
\lref\becker{K. Becker, M. Becker and A. Strominger, hep-th/9507158,
\NP {\bf B456} (1995) 130.}
\lref\vafa{M. Bershadsky, V. Sadov and C. Vafa, hep-th/9511222, \NP
{\bf B463} (1996) 420.}
\lref\gaunt{J.P. Gauntlett and D.A. Lowe, hep-th/9601085;
S.A. Connell, unpublished; K. Lee, E.J. Weinberg and P. Yi,
hep-th/9601097, hep-th/9602167; G.W. Gibbons, hep-th/9603176.}
\lref\klv{S. Kachru, A. Klemm, W. Lerche, P. Mayr and C. Vafa,
hep-th/9508155, \NP {\bf B459} (1996) 537.}
\lref\hitchin{N.J. Hitchin, {\it Math. Proc. Camb. Phil. Soc.}
{\bf 85} (1979) 465.}
\lref\ddvafa{M. Bershadsky, V. Sadov and C. Vafa, hep-th/9510225,
\NP {\bf B463} (1996) 398.}
\lref\bikm{M. Bershadsky, K. Intriligator, S. Kachru, D.R. Morrison, V.
Sadov and C. Vafa, hep-th/9605200.}
\lref\oog{H. Ooguri and C. Vafa, hep-th/9511164, \NP {\bf B463} (1996) 55.}
\lref\doug{M.R. Douglas and G. Moore, hep-th/9603167.}
\lref\katz{S. Katz, D.R. Morrison and M.R. Plesser, hep-th/9601108.}
\lref\klmvw{A. Klemm, W. Lerche, P. Mayr, C. Vafa
and N. Warner, hep-th/9604034.}
\lref\aspinb{P.S. Aspinwall, hep-th/9511171, \PL {\bf B371} (1996) 231.}
\lref\sw{N. Seiberg and E. Witten, hep-th/9408099, \NP {\bf B431} (1994) 484.}
\lref\dw{R. Donagi and E. Witten, hep-th/9510101, \NP {\bf B460} (1996) 299.}
\lref\hullt{C.M. Hull and P.K. Townsend, hep-th/9410167,
\NP {\bf B438} (1995) 109.}
\lref\porrati{M. Porrati,  hep-th/9505187.}

%
\baselineskip 12pt
\Title{\vbox{\baselineskip12pt
\line{\hfil  UCSBTH-96-14}
\line{\hfil  NSF-ITP-96-58}
\line{\hfil \tt hep-th/9606146} }}
{\vbox{
{\centerline{Evidence for S-Duality in}}
{\centerline{N=4
Supersymmetric Gauge Theory}}
}}
\centerline{\ticp Karl Landsteiner$^\dagger$,\footnote{}{\ttsmall
karll@cosmic1.physics.ucsb.edu, elopez@itp.ucsb.edu,
lowe@tpau.physics.ucsb.edu
} Esperanza L\'opez$^\ddagger$ and David A. Lowe$^\dagger$}
\bigskip
\vskip.1in
\centerline{$^\dagger$\it Department of Physics, University of California,
Santa Barbara, CA 93106, USA}
\vskip.1in
\centerline{$^\ddagger$\it Institute for Theoretical Physics,
University of California,
Santa Barbara, CA 93106, USA}
\bigskip
\centerline{\bf Abstract}
Using D-brane techniques, we compute the spectrum of stable
BPS states in N=4 supersymmetric gauge theory, and
find it is consistent with Montonen-Olive duality.

\Date{June, 1996}

Gauge theory in four dimensions with $N=4$
supersymmetry has been conjectured by
Montonen and Olive to possess
a strong/weak coupling duality symmetry \refs{\mont,\osborn}.
For the simplest case, with $SU(2)$ gauge symmetry,
the duality group is $SL(2,\IZ)$. By considering
the orbits of the states in the $N=4$ gauge
multiplet under this $SL(2,\IZ)$, a whole tower
of BPS\footnote{$^\#$}{In this paper we will reserve
the term BPS for states which preserve one half of the supersymmetry.}
monopole and dyon states are predicted
with magnetic-electric charges given by $(p,q)$ with $p$ and $q$
relatively prime \sen. Proving the
existence of these states
is an important check of the Montonen-Olive S-duality conjecture.

The states with magnetic charge one are the usual
't Hooft-Polyakov monopole \hooft\ and the
Julia-Zee dyons \zee\ embedded in $N=4$ gauge theory. Sen showed
the states with magnetic charge two (and odd electric charge)
arise as bound states of two singly charged monopoles \sen.
States with magnetic charge larger than two will arise as
bound states of multi-monopole configurations.
The construction of these states as bound states in the
supersymmetric quantum mechanics on the moduli space of these
configurations remains an open problem (see
\porrati\ for results in this direction).
In this paper, we will take a different approach and
use the recently developed D-brane technology \joe\ to prove the
existence of the whole tower of $(p,q)$-states with
degeneracies in accord with S-duality. We also show that
this generalizes in a straightforward way to $ADE$ groups.

The starting point for this construction is the
Type IIB string theory compactified on $K3\times T^2$ \refs{\hullt \witten
\ddvafa {--} \katz}.
This gives rise to a theory in four dimensions
with $N=4$ supersymmetry. One considers a point in
the $K3$ moduli space near which the Type IIA string
develops an enhanced gauge symmetry \refs{\witdy, \aspin}. In this
limit the $K3$ develops an orbifold singularity of $ADE$ type as
a set of two-spheres shrink to zero size.
Type II string theory compactified on this and closely related
manifolds have recently been considered in \refs{\klv \klmvw \bikm{--}
\aspinb}, and D-branes on such manifolds have been
studied in \refs{\ddvafa, \klmvw, \oog, \doug}.
{}From the Type IIB
perspective the relevant BPS states arise from
self-dual supersymmetric
threebranes \duff\ wrapping the supersymmetric three-cycles \becker\ of
$K3\times T^2$ which are shrinking to zero size.
These three-cycles are
the product of the two-sphere (which is
a holomorphic two-cycle of the $K3$ with respect to an
appropriate choice of complex structure) with
a one-cycle of the torus.
In string units the mass of such a state is
\eqn\thmass{
M \sim {\epsilon R \over \lambda}~,
}
where $\epsilon$ is the area of the $S^2$, $R$ is the size of the
one-cycle of the torus and $\lambda$ is the Type IIB string
coupling constant. Since $\epsilon$ can be made arbitrarily small
by moving toward the orbifold singularity of the $K3$, we may
consider a limit in which gravity, the other fundamental IIB
string excitations, and the Kaluza-Klein modes, are irrelevant.
The effective four-dimensional theory describing these
threebrane states will be some $N=4$ gauge theory \witten, valid
up to some cutoff $\Lambda = {\rm min}(1/R,
\sqrt{\epsilon/\lambda})$.\footnote{$^*$}{This
cutoff could be removed by
going to the point-particle limit along the lines of \klv (also see final
paragraph).}

Let us first consider the case when a single two-sphere shrinks
to zero size, corresponding to $SU(2)$ gauge theory.
The $U(1)$ gauge field $A$ of this theory arises from the
4-form potential $C$ of the IIB theory
\eqn\gfield{
C= A \wedge G \wedge dx~,
}
where $G$ is a self-dual harmonic two-form on $K3$ with support on the
neighborhood of the $S^2$,
and $z=x+iy$ is the complex
coordinate of $T^2$, with periodic identification
$z\sim z+1$, $z\sim z+\tau$. Note that because $C$ is the
potential for a self-dual field strength, only one independent
$U(1)$ gauge field appears in four dimensions.

We may regard the states wrapping an A-cycle (i.e. in the $x$-direction)
of $T^2$ $q$-times and the B-cycle $p$-times
as carrying magnetic and electric charge $(p,q)$.
The BPS condition requires the three-cycle have minimal volume, which
in turn implies that
these $(p,q)$-cycles are straight lines in the $z$-plane. The masses
of these states are therefore
\eqn\smass{
M = {\epsilon R\over \lambda} |p + \tau q|~.
}
We see
the coupling constant of the gauge theory is to
be identified with the complex structure parameter $\tau$ of the
torus. Note the $SL(2,\IZ)$ T-duality symmetry of the Type IIB
which acts on $\tau$ is mapped to the S-duality symmetry of
the four-dimensional gauge theory. Proving T-duality of the
nonperturbatively defined IIB string theory would then
amount to proving S-duality of $N=4$ gauge theory.

To complete this check of the S-duality conjecture we must
compute the degeneracy of these $(p,q)$-states.
To accomplish this we will move to a point in the
moduli space of the IIB theory at which the three-cycle becomes
very large, so that the mass of the solitonic states becomes
enormous compared to the masses of the fundamental string
excitations.
Since we have $N=4$ supersymmetry, we know the
degeneracy of these states will not change as we move
in the moduli space.
Type IIB string
perturbation theory will be valid in this limit, and we may use the D-brane
picture
\joe\ to describe these threebrane states.

In this limit we can use the $R \rightarrow 1/R$
duality of perturbative string theory acting on the circle of
the two-torus which is defined by $(p,q)$. This will also map
type IIB to type IIA theory and three-branes to two-branes.
If the greatest common divisor of $p$ and $q$ is $n$, then
the corresponding state will be mapped to $n$ two-branes of
the type IIA theory wrapping the $S^2$ of the $K3$.
In this situation the number of BPS states which
preserve one half the supersymmetry, provided by
D-brane configurations, is a topological invariant (related
to the dimension of the cohomology of
the moduli space of the corresponding
Hitchin system \vafa). For $n>1$ we may regard the $n$ D-branes
wrapping the $S^2$ as $n$ copies of the $S^2$, i.e. as
a two-cycle with self-intersection number $-2n^2$ \vafa. It may
be shown that such a cycle can not be realized as a smooth
holomorphic curve, hence no supersymmetric bound states will
be produced. For $n=1$, the $S^2$ can always be realized as
a smooth genus zero holomorphic curve and the moduli space
of the relevant Hitchin system is a point. The dimension
of the cohomology is one, so a single supersymmetric bound state
exists.
This completes the argument.

These results for $SU(2)$ extend in a straightforward
way to higher rank gauge theories of $ADE$
type.\footnote{$^\flat$}{Note that since the group must
appear as an orbifold singularity of $K3$ we are restricted
to groups with rank $\leq 19$ (or $20$ if we include nontrivial $B$-fields).}
S-duality predicts a single BPS multiplet with magnetic-electric
charges $(p,q)\valpha$ for each root $\valpha$ of the group,
with $p$,$q$ relatively prime.
In the D-brane picture, these will arise from
threebranes partially wrapping two-spheres shrinking to
zero size as an orbifold singularity (of $A$, $D$ or $E$ type)
of the $K3$ is approached.

For our purposes it is sufficient
to study the behavior of the geometry near the singularity,
which allows us to replace the $K3$ by an ALE space ${\cal M}$ of type
$G$ \hitchin.
The homology $H_2(\M, \IZ)\cong \IZ^r$ ($r$ is the rank of $G$)
and the intersection
form is $-C_G$, the Cartan matrix for the
group $G$, thus the integral homology is identified
with the root lattice of the group. The cohomology group
$H^2(\M,\IZ)$ is identified with the weight lattice of $G$ and
corresponds to a set of anti-self-dual two-forms. In addition,
there are three covariantly constant self-dual two-forms $\vec \omega$
corresponding to the K\"ahler form and the holomorphic
two-form (with respect to some choice of complex structure). These self-dual
forms are
non-normalizable on the ALE space.
We denote the periods of these three two-forms by
\eqn\periods{
\int_{\Sigma_i} \vec \omega = \vec \zeta_i~,
}
for some choice of basis $\Sigma_i$ ($i=1,\cdots,r$) of $H_2(\M,\IZ)$,
corresponding to a choice of simple roots. The area
of a two-cycle corresponding to a root $\valpha$ will be
\eqn\tsize{
\epsilon \propto |\valpha \cdot \vec \zeta|~.
}

When we consider the dyon states associated with a simple
root, the D-branes will wrap a single two-sphere
$\Sigma_i$ and the above results for the $SU(2)$ case will
go through. For states associated with nonsimple root $\valpha$, the
D-branes will  wrap the product of a collection of intersecting
two-spheres with a single $S^1$ of the two-torus. For generic values of
the moduli of the ALE space
we may nevertheless regard this collection of intersecting two-spheres
as a single two-sphere. This may be
realized as a smooth holomorphic genus zero curve, and
we may therefore again apply the preceding arguments which show
that
one BPS multiplet with charge $(p,q)\valpha$ exists for every
$p$,$q$ relatively prime.
Recalling that the moduli of the ALE space are
simply the $\vec \zeta_i$ modulo the Euclidean group
in three dimensions, one sees from \tsize\ that
for generic values of the moduli
these states will be true bound states.
However
a subtlety appears when we adjust the moduli so that two or
more of the $\vec \zeta_i$ are parallel. At this point a
nonsimple root state may be only neutrally stable with
respect to decay into simple root states.
In this limit, the corresponding two-cycle in the D-brane
picture will become degenerate, corresponding
to two two-spheres intersecting at a point, for example. Using the fact that no
jumping phenomena can occur in theories with $N=4$
supersymmetry, we may nevertheless argue that a bound state
at threshold must appear.

This agrees with recent field theory results which
construct these states at threshold as bound states
in the supersymmetric quantum mechanics on the moduli
space of monopoles of distinct type \refs{\gaunt}. The moduli
spaces in question were constructed for a single real Higgs
field. We expect that for generic asymptotic vevs of two or more
Higgs fields,\footnote{$^\natural$}{With $N=4$ supersymmetry
we have a total of six adjoint Higgs fields present.
These give rise to moduli corresponding to the moduli of the ALE
space, with B-field, together with additional moduli coming from RR
fields. Note that only
a subset of these moduli (related to the $\vec \zeta_i$) show up in our
present construction, based
on classical geometry.}
the analogous moduli spaces will collapse
to a simple $R^3 \times S^1$ form.

We must also show no other bound states arise in the
D-brane picture. Clearly
no new bound states will come from D-threebranes wrapping disconnected sets of
two-spheres,
so we need only consider a connected set corresponding
to some point on the root lattice. In order to get a
supersymmetric configuration, it is necessary to take the product of
the two-spheres with parallel $S^1$'s on the torus, as opposed to different
nonparallel $S^1$'s for different two-spheres.
One may then
move to a point in moduli space where IIB perturbation theory
applies and apply $R\to 1/R$ duality in the
$S^1$ direction defined by $(p,q)$
(with $p$,$q$ relatively prime).  This yields
a configuration
equivalent to a single D-twobrane wrapping some
element $\Sigma \in H_2(\M,\IZ)$.
This element will have self-intersection number
$\Sigma^2 \leq -2$.
However such an element may only be realized as a holomorphic
two-cycle when $\Sigma^2 \geq -2$, thus only the roots of the
gauge group (which are in one-to-one correspondence with
the $\Sigma^2=-2$ elements) give rise to supersymmetric bound states, as
described above.

Finally, let us notice that by going to the point particle limit
of the string \klv, it is possible to obtain a rigid field theory
valid for all energy scales. The point particle limit fixes the
$K3$ moduli parameters at the values where it develops the $ADE$
orbifold point, and resolves this singularity by blowing up the
singular $K3$ to an ALE space $\cal M$. We can fix the K\"ahler form of the
ALE space in such a way that the Higgs fields $\phi$ take only
values in a two-dimensional subspace. In this situation the $\vec \zeta_i$
are restricted to move in an $R^2$ subspace and their configuration
is isomorphic to that defined by the set of points over the complex plane
\eqn\periods{
V= \{ x \vert P_G= {\rm det}(\phi -x)=0 \}.
}
Therefore the information contained in $\M \times T^2$
is alternatively encoded in the curve defined by the
trivial fibration of the set $V$ over the torus $T^2$. This curve
coincides with the one proposed in \refs{\sw,\dw} for representing the
moduli space of $N=4$ Yang-Mills theory with gauge group $G$.
In the same way as for $N=2$ string theories \klmvw\ we see that for
$N=4$ it is also possible to recover from the string compactification,
in addition to the effective gauge theory physics,
its associated Seiberg-Witten curve.

\vskip 1cm
{\bf Acknowledgements}

\vskip .5cm

The research of D.L is
supported in part by NSF
Grant PHY91-16964, that of K.L. by the Fonds zur F\"orderung der
wissenschaftlichen Forschung under Grant J01157-PHY and that of E.L. by
a C.A.P.V. fellowship.

\listrefs
\end